\documentclass[aps,prd,onecolumn]{revtex4-1}
\usepackage{epsfig}
\usepackage{color}

\begin{document}
\author{Pei-lin Yin$^{1}$, Zhu-fang Cui$^{2,4}$, Hong-tao Feng $^{3,4}$, and Hong-shi Zong$^{2,4,5}$}\email[]{zonghs@chenwang.nju.edu.cn}
\address{$^{1}$Key Laboratory of Modern Acoustics, MOE, Institute of Acoustics, and Department of Physics, Nanjing University, Nanjing 210093, China}
\address{$^{2}$Department of Physics, Nanjing University, Nanjing 210093, China}
\address{$^{3}$Department of Physics, Southeast University, Nanjing 211189, China}
\address{$^{4}$State Key Laboratory of Theoretical Physics, Institute of Theoretical Physics, CAS, Beijing, 100190, China}
\address{$^{5}$Joint Center for Particle, Nuclear Physics and Cosmology, Nanjing 210093, China}

\title{The chiral phase transition of QED$_3$ around the critical number of fermion flavors}

\begin{abstract}
At zero temperature and density, the nature of the chiral phase transition in QED$_3$ with $\textit{N}_{f}$ massless fermion flavors is investigated. To this end, in Landau gauge, we numerically solve the coupled Dyson-Schwinger equations for the fermion and boson propagator within the bare and simplified Ball-Chiu vertices separately. It is found that, in the bare vertex approximation, the system undergoes a high-order continuous phase transition from the Nambu-Goldstone phase into the Wigner phase when the number of fermion flavors $\textit{N}_{f}$ reaches the critical number $\textit{N}_{f,c}$, while the system exhibits a typical characteristic of second-order phase transition for the simplified Ball-Chiu vertex.
\bigskip

\noindent Keywords: QED$_3$, Chiral phase transition, Dyson-Schwinger equations
\bigskip

\noindent PACS Number(s):  11.10.Kk, 11.15.Tk, 11.30.Qc
\end{abstract}
\maketitle

\section{Introduction}
Dynamical chiral symmetry breaking (DCSB) and confinement are two fundamental features of Quantum Chromodynamics (QCD). Studying these two aspects, will provide profound insight into the origin of observable mass and the nature of the early Universe. However, due to the complicated non-Abelian character of QCD, it is difficult to have a thorough understanding of the mechanism of DCSB and confinement. In this case, to gain valuable comprehension about them, it is suggested to study some model which is similar to QCD and, at the same time, simpler. Quantum Electrodynamics in (2+1) dimensions (QED$_3$) is such a model, and it has been studied quite intensively over the past few years~\cite{PhysRevD.29.2423,PhysRevLett.55.1715,PhysRevLett.60.2575,PhysRevLett.62.3024,PhysRevD.44.540,PhysLettB.295.313,PhysRevLett.74.18,PhysRevD.54.4049,
PhysLettB.491.280,PhysRevD.68.025017,PhysRevD.73.016004,PhysRevC.78.055201,PhysRevD.84.036013,PhysRevD.86.105042,CommunMathPhys.82.545,PhysRevD.46.2695,PhysRevD.52.6087,
PhysRevLett.91.171601}. In addition, due to the coupling constant being dimensionful (its dimension is $\sqrt{\rm{mass}}$), QED$_3$ is super-renormalizable, so it does not suffer from the ultraviolet divergences which are present in QED$_4$. Apart from these interesting features, QED$_3$ with $\textit{N}_{f}$ massless fermion flavors can also be regarded as a possible low energy effective theory for strongly correlated electronic systems~\cite{PhysRevLett.86.3871,PhysRevB.66.054535,
PhysRevB.71.172501,RevModPhys.78.17,PhysRevLett.102.026802,PhysRevB.83.115438}.

In the last few decades, whether a critical number of fermion flavors $\textit{N}_{f,c}$, where the system undergoes a phase transition from the Nambu-Goldstone phase (NG phase) into the Wigner phase (WN phase), exists or not has been discussed intensively. A breakthrough was achieved in 1988 by T.W. Appelquist \textit{et al.}~\cite{PhysRevLett.60.2575}. Using the bare vertex approximation and the one-loop vacuum polarization, they solved the Dyson-Schwinger equations (DSEs) for the fermion self-energy and found that chiral phase transition occurs when the number of fermion flavors reaches a critical number $\textit{N}_{f,c}$= $32/\pi^{2}$. Subsequently, P. Maris solved the coupled DSEs for the fermion self-energy and boson vacuum polarization with a range of simplified fermion-boson vertices and obtained a critical number of fermion flavors $\textit{N}_{f,c}$= 3.3 ~\cite{PhysRevD.54.4049}. Recently, A. Bashir \textit{et~al}. analyzed the characters of fermion wave function renormalization and boson vacuum polarization at the infrared momenta when the fermion mass function vanishes and arrived at $\textit{N}_{f,c}$ $\approx$ 3.24 employing a model for the photon vacuum polarization and fermion每boson vertex ~\cite{PhysRevC.78.055201}.

As far as we know, however, in the existing literature the studies of the nature of chiral phase transition around the critical number of fermion flavors $\textit{N}_{f,c}$ are relatively scarce. The authors of Ref.~\cite{PhysRevD.86.105042} discussed this question by solving DSEs for the fermion self-energy in the lowest-order approximation and drew the conclusion that the chiral phase transition in QED$_3$ with $\textit{N}_{f}$ massless fermion flavors is a continuous phase transition higher than second-order. In the present paper, we try to reanalyze the nature of this phase transition by numerically solving the coupled DSEs for the fermion and boson propagator within the bare and simplified Ball-Chiu vertices separately at zero temperature and density.

This paper is organized as follows. In Section II, we firstly introduce the criteria determining the locations and characteristics of the chiral phase transition, and then derive the DSEs satisfied by the fermion and boson propagator of QED$_3$. In Section III, we discuss the behaviors of the fermion and boson propagator in NG phase and WN phase by numerically solving the coupled DSEs in the truncated scheme and then study how the chiral condensate, the chiral susceptibility, the differential pressure between these two phases, and the infrared values of self-energy function change with the variation of the number of fermion flavors. In Section IV, we will briefly summarize our results and give the conclusions.
\section{chiral phase transition and DSEs in QED$_3$}
In quantum field theory, the dynamic properties of a system are fully characterized by the generating functional corresponding to the partition function in thermodynamics. It is commonly accepted that when the system is in a certain phase, the generating functional is usually analytic for some choice of parameters, such as the current mass of the fermion, the temperature and the chemical potential; the generating functional often exhibits the non-analytic character while the phase transitions occur. So the location and characteristic of the chiral phase transition in the system can be determined by the behaviors of this quantity with respect to the corresponding parameters (i.e., the current mass, the temperature and the chemical potential). In this case, a phase transition, in which the first-order derivative of the generating functional with some of the parameters is discontinuous, is referred to as first-order or discontinuous phase transition. Second-order or continuous phase transition exhibits the continuity in first-order derivative and the discontinuity or infinity in second-order derivative.

\subsection{Criteria for chiral phase transition}
The chiral condensate is the vacuum expectation value for the scalar operator$\bar\psi\psi$. The character, the nonzero value of it indicates that chiral symmetry reflected on the Lagrangian level is spontaneously broken on the vacuum level and the chiral symmetry gets restored when the chiral condensate vanishes for the chiral limit, makes it possible to define the chiral condensate as the order parameter for the chiral phase transition. The chiral condensate is commonly given by the first-order derivative of the generating functional with respect to the current mass of the fermion
\begin{eqnarray}
\langle\bar\psi\psi\rangle=-\frac{\partial LnZ}{\partial m}= -\int\frac{\textrm{d}^3p}{(2\pi)^3}\textrm{Tr}[S(p)],\label{eq1}
\end{eqnarray}
where S is the dressed fermion propagator and the notation Tr denotes trace operation over Dirac indices of the fermion propagator.

In addition to the above chiral condensate, the susceptibility representing a response of the system to an external perturbation is often investigated. The feature, the susceptibility usually exhibits some of singular behaviors, such as discontinuity or infinity, when the chiral phase transition occurs, enables it to be used for studying the chiral phase transition ~\cite{Nature.443.675,PhysRevD.82.054026}. Herein we consider the chiral susceptibility that can be written as the first-order derivative of chiral condensate with respect to the current mass of the fermion
\begin{eqnarray}
\chi_{c}=\frac{\partial(-\langle\bar\psi\psi\rangle_{m})}{\partial m}\bigg|_{m\rightarrow0}=\frac{\partial^{2} LnZ}{\partial m^{2}}\bigg|_{m\rightarrow0}
=-\int\frac{\textrm{d}^3p}{(2\pi)^3}\textrm{Tr}[S(m,p)\frac{\partial \textit {S}^{-1}(m,p)}{\partial m}S(m,p)]\bigg|_{m\rightarrow0},\label{eq2}\end{eqnarray}
This equation indicates that the chiral susceptibility measures the response of the chiral condensate to an infinitesimal change of the fermion mass.

Furthermore, in order to analyze the effect of the number of fermion flavors on the chiral phase transition more directly, we need to calculate the derivative of the generating functional with respect to the number of fermion flavors. Because of the character of non-perturbation, we cannot obtain an exact expression for the generating functional. However, in some situations, an expression for the effective potential can be given in terms of the fermion and boson propagator, which permits us to carry out the calculation of the derivative. In the present paper, we adopt the CJT effective potential~\cite{PhysRevD.10.2428}, which corresponds to the bare vertex approximation for solving the DSEs for the fermion and the boson propagator, to study the nature of the chiral phase transition around the critical number of fermion flavors. The pressure is the negative of the CJT effective potential density. At zero temperature and density the effective pressure is given as
\begin{eqnarray}
P(N_{f})=-\textit{N}_{f}\times\textrm{Tr[Ln}(S_{0}^{-1}S)+\frac{1}{2}\textrm(1-S_{0}^{-1}S) ]+\frac{1}{2}\times\textrm{Tr[Ln}(D_{0}^{-1}D)+(1-D_{0}^{-1}D) ],\label{eq3}
\end{eqnarray}
where the trace, the logarithm, and the product of propagator are taken in the functional sense. The first item represents, the contribution of $\textit{N}_{f}$ massless fermion flavors, and the last item denotes the contribution of photon, which is rarely discussed in the existing literature. Because of the divergent integral, the differential pressure between the NG phase and WN phase is often calculated
\begin{eqnarray}
\Delta P(N_{f})=P_{\it {NG}}(\textit{N}_{f})-P_{\it {WN}}(\textit{N}_{f}),\label{eq4}
\end{eqnarray}

To determine the order of the chiral phase transition that we concern, herein we regard $\Delta P(N_{f})$ as a continuous function of $\textit{N}_{f}$. More specifically, we can expand $\Delta P(N_{f})$ by the Taylor series near the critical number of fermion flavors $\textit{N}_{f,c}$
\begin{eqnarray}
\Delta P(N_{f})= \Delta P(N_{f,c}) + (\textit{N}_{f}-\textit{N}_{f,c})\frac{\partial(\Delta P(N_{f}))}{\partial\textit{N}_{f}}\bigg|_{{N_{f}}=\it {N}_{f,c}}+\frac{(\textit{N}_{f}-\textit{N}_{f,c})^{2}}{2}\frac{\partial^{2}(\Delta P(\textit{N}_{f}))}{\partial\textit{N}_{f}^{2}}\bigg|_{{N_{f}}=\it {N}_{f,c}}+\cdot\cdot\cdot,\label{eq5}
\end{eqnarray}
with
\begin{eqnarray}
\Delta P'(\textit{N}_{f})=\frac{\partial(\Delta P(N_{f}))}{\partial\textit{N}_{f}},\label{eq6}
\end{eqnarray}

\begin{eqnarray}
\Delta P''(\textit{N}_{f})=\frac{\partial^{2}(\Delta P(\textit{N}_{f}))}{\partial\textit{N}_{f}^{2}},\label{eq7}
\end{eqnarray}
The phase transition is a first-order one when $\Delta P'(\textit{N}_{f})$ is discontinuous at $\textit{N}_{f}=\textit{N}_{f,c}$ and its continuity might imply a second-order phase transition.
\subsection{Dyson-Schwinger equations in QED$_3$}
According to the fundamental theory of group representations, the dimension of a spinorial representation for the Lorentz group must be even. In (2+1) dimension only three $\gamma$ matrices that satisfy the corresponding Clifford algebra are needed, meanwhile, in quantum mechanics there have been three anticommutative matrices that are just 2$\times$2 Pauli spin matrices, so the two-dimensional matrix representation and two-component spinors are sufficient. However, there is no other 2$\times$2 matrix that anticommutes with all three $\gamma$ matrices. There is, therefore, nothing to generate a chiral symmetry and so we cannot discuss the chiral symmetry. Besides, the possible mass term has the undesirable property that it is odd under the parity transformation. Given this, we employ the fourdimensional matrix representation and four-component spinors as in four space每time dimensions in this paper. 

In Euclidean space, the Lagrangian density of QED$_3$ with $\textit{N}_{f}$ massless fermion flavors reads
\begin{equation}
\mathcal{L}=\sum _{i=1}^N \bar{\psi }_i({\not\!\partial} +ie{\not\!A})\psi _i+\frac{1}{4}F_{\mu \nu }^2+\frac{1}{2\xi }\left(\partial _{\mu }A_{\mu }\right){}^2,\label{eq8}
\end{equation}
where the spinor $\psi_i$ is the fermion field with the indices \textit{i}=1,...,$\textit{N}_{f}$ representing different fermion flavors, $A_{\mu }$ is the electromagnetic vector potential, $F_{\mu \nu }$ is the electromagnetic field strength tensor, and $\xi$ represents the gauge parameter (we will adopt the Landau gauge $\xi=0$ throughout this paper). Using this Lagrangian density one can derive in the standard way, for instance through functional analysis, the DSEs for the propagators.

For the fermion propagator the DSEs can be written as
\begin{equation}
S^{-1}(p)=S_{0}^{-1}(p) + \Sigma(p),\label{eq9}
\end{equation}
\begin{equation}
\Sigma(p)=\int\frac{\textrm{d}^{3}k}{(2\pi)^{3}}\gamma_{\mu}S(k)\Gamma_{\nu}(p,k)D_{\mu\nu}(q),\label{eq10}
\end{equation}
where $S(p)$ and $S_0(p)=1/{i\gamma\cdot p}$ are the dressed fermion propagator and the free fermion propagator in the chiral limit, respectively, $\Sigma(p)$ is the fermion self-energy, $\Gamma_{\nu}(p,k)$ is the full fermion每boson vertex, and $D_{\mu\nu}(q)$ is the dressed photon propagator. Meanwhile, based on Lorentz structure analysis, the fermion propagator can be written as
\begin{equation}
S^{-1}(p)=i{\not\!p}A(p^2)+B(p^2),\label{eq11}
\end{equation}
where both $A(p^2)$ and $B(p^2)$ are scalar functions of $p^2$. Substituting Eq. (\ref{eq10}) and Eq. (\ref{eq11}) into Eq.(\ref{eq9}), one can immediately obtain
\begin{equation}
A(p^2)=1-\frac{i}{4p^{2}}\int\frac{\textrm{d}^{3}k}{(2\pi)^{3}} \textrm{Tr}[\gamma\cdot p \gamma_{\mu}S(k)\Gamma_{\nu}(p,k)D_{\mu\nu}(q)],\label{eq12}
\end{equation}
\begin{equation}
B(p^2)=\frac{1}{4}\int\frac{\textrm{d}^{3}k}{(2\pi)^{3}} \textrm{Tr}[\gamma_{\mu}S(k)\Gamma_{\nu}(p,k)D_{\mu\nu}(q)],\label{eq13}
\end{equation}

The DSEs for the photon propagator have the form
\begin{equation}
D_{\mu\nu}^{-1}(q)=D_{\mu\nu}^{0,-1}(q) + \Pi_{\mu\nu}(q),\label{eq14}
\end{equation}
\begin{equation}
\Pi_{\mu\nu}(q)= -N_{f}\int\frac{\textrm{d}^{3}k}{(2\pi)^{3}}\textrm{Tr}[\gamma_{\mu}S(k)\Gamma_{\nu}(p,k)S(p)],\label{eq15}
\end{equation}
where $D_{\mu\nu}^{0}(q)=(\delta_{\mu\nu}-q_{\mu}q_{\nu}/q^{2})/q^{2}$ is the free photon propagator,
$\Pi_{\mu\nu}(q)$ is the vacuum polarization tensor. At the same time, in order to ensure the Ward-Takahashi identity, $\Pi_{\mu\nu}(q)$ has the form
\begin{equation}
\Pi_{\mu\nu}(q)=(q^{2}\delta_{\mu\nu}-q_{\mu}q_{\nu})\Pi(q^{2}),\label{eq16}
\end{equation}
where $\Pi(q^{2})$ is the photon self-energy, i.e., the vacuum polarization. Substituting Eq. (\ref{eq16}) into Eq. (\ref{eq14}), one has
\begin{equation}
D_{\mu\nu}(q)=\frac{(\delta_{\mu\nu}-q_{\mu}q_{\nu}/q^{2})}{q^{2}(1+\Pi(q^{2}))},\label{eq17}
\end{equation}
However, we also note that the vacuum polarization tensor $\Pi_{\mu\nu}(q)$ has an ultraviolet divergence that is present only in the longitudinal part. By applying the following projection operator ~\cite{PhysRevD.46.2695}:
\begin{equation}
\mathcal{P_{\mu\nu}}=\delta_{\mu\nu}-3\frac{q_{\mu}q_{\nu}}{q^{2}},\label{eq18}
\end{equation}
one can remove this divergence and project out a finite vacuum polarization $\Pi(q^{2})$,
\begin{equation}
\Pi(q^{2})=\frac{\delta_{\mu\nu}-3q_{\mu}q_{\nu}/q^{2}}{2q^{2}}\Pi_{\mu\nu}(q), \label{eq19}
\end{equation}

From Eq. (\ref{eq12}), Eq. (\ref{eq13}), Eq. (\ref{eq17}), Eq. (\ref{eq19}), and Eq. (\ref{eq15}), one can see that if the full fermion-boson vertex $\Gamma_{\nu}(p,k)$ is known, the fermion and boson gap equations form a set of equations that can be solved numerically by the iteration method. In the past few years, there are several attempts to determine the form of $\Gamma_{\nu}(p,k)$ in the literature~\cite{PhysRevD.44.540,PhysLettB.295.313,PhysRevLett.74.18}. It should be mentioned here that the authors of Refs.~\cite{PhysRevC.85.045205,PhysLettB.722.384} analyzed the general Lorentz structure of dressed fermion每photon vertex in the context of considering the constraints of the gauge symmetry and presented a workable model for the dressed fermion每photon vertex. In the present paper, following the Ref.~\cite{PhysRevD.54.4049}, we choose the following Ans\"{a}tze for the fermion每photon vertex
\begin{equation}
\Gamma_{\nu}(p,k)=f(A(p^{2}),A(k^{2}))\gamma_{\nu},\label{eq20}
\end{equation}
and the form of $f(A(p^{2}),A(k^{2}))$ is: (1) 1; (2)~ $\frac{A(p^{2})+A(k^{2})}{2}$. The first one is the bare vertex and plays the most dominant role in the large momentum limit. The second form is inspired by the Ball-Chiu (BC) vertex. Previous works show that the numerical results of DSEs obtained employing this choice are in
good agreement with the results obtained employing the BC and Curtis-Pennington (CP) vertices, so we choose this one to be used in following calculation.

Substituting Eq. (\ref{eq17}) into Eq. (\ref{eq12}) and Eq. (\ref{eq13}), and similarly substituting Eq. (\ref{eq11}) and Eq. (\ref{eq15}) into Eq. (\ref{eq19}), we can write down the coupled fermion and boson gap equations in the following form:
\begin{eqnarray}
A(p^2)=1+\frac{2}{p^{2}}\int\frac{\textrm{d}^{3}k}{(2\pi)^{3}}\frac{A(k^2)}{A^{2}(k^2)k^2+B^{2}(k^2)}\frac{(p\cdot q)(k\cdot q)/q^{2}
f(A(p^{2}),A(k^{2}))}{q^{2}(1+\Pi(q^{2}))},\label{eq21}
\end{eqnarray}
\begin{equation}
B(p^2)=2\int\frac{\textrm{d}^{3}k}{(2\pi)^{3}}\frac{B(k^2)}{A^{2}(k^2)k^2+B^{2}(k^2)}\frac{f(A(p^{2}),A(k^{2}))}{q^{2}(1+\Pi(q^{2}))},\label{eq22}
\end{equation}
\begin{eqnarray}
\Pi(q^2)=\frac{4N_{f}}{q^{2}}\int\frac{\textrm{d}^{3}k}{(2\pi)^{3}}\frac{A(k^2)}{A^{2}(k^2)k^2+B^{2}(k^2)}\frac{A(p^2)}{A^{2}(p^2)p^2+B^{2}(p^2)}(k^{2}-2(k\cdot q)-3(k\cdot q)^{2}/q^{2})f(A(p^{2}),A(k^{2})),\label{eq23}
\end{eqnarray}

\section{numerical results}
\subsection{The behavior of the propagator in the two phases}
\begin{figure}[h]
\begin{minipage}[t]{0.47\textwidth}
\includegraphics[width=\textwidth]{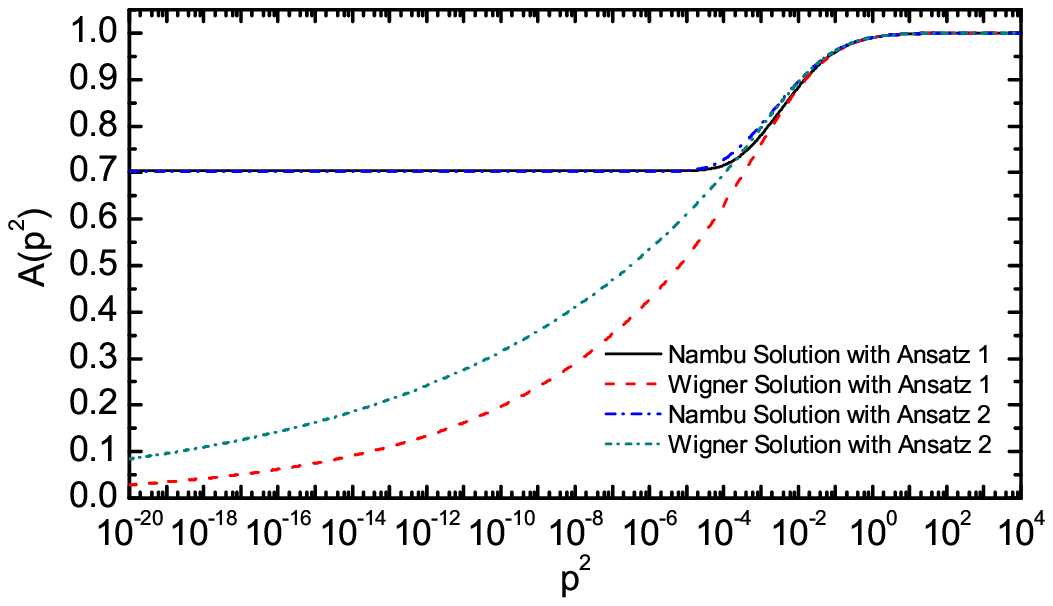}
\end{minipage}
\hfill
\begin{minipage}[t]{0.47\textwidth}
\includegraphics[width=\textwidth]{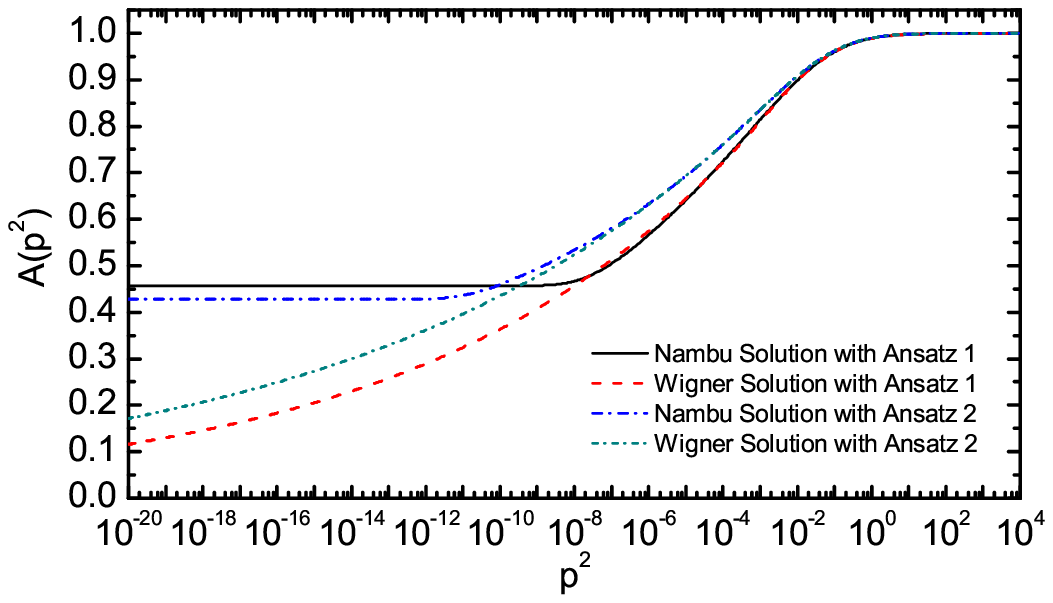}
\end{minipage}
\caption{The behavior of $A(p^{2})$ with the variation of $p^{2}$ for $\textit{N}_{f} =2$ (left) and $\textit{N}_{f} =3$ (right).}\label{fig1}
\end{figure}

\begin{figure}[h]
\begin{minipage}[t]{0.47\textwidth}
\includegraphics[width=\textwidth]{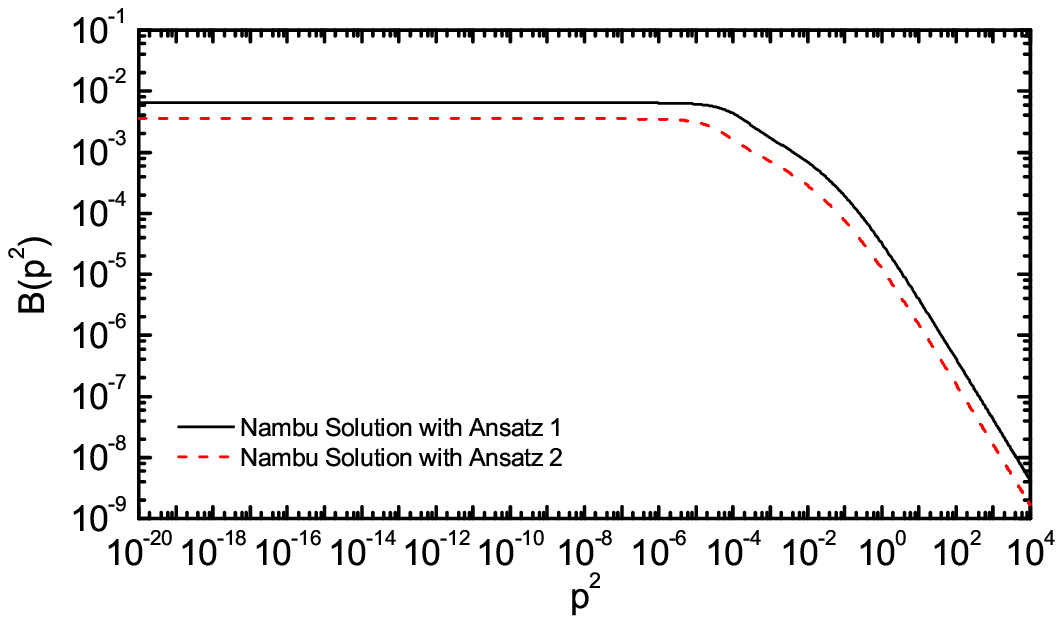}
\end{minipage}
\hfill
\begin{minipage}[t]{0.47\textwidth}
\includegraphics[width=\textwidth]{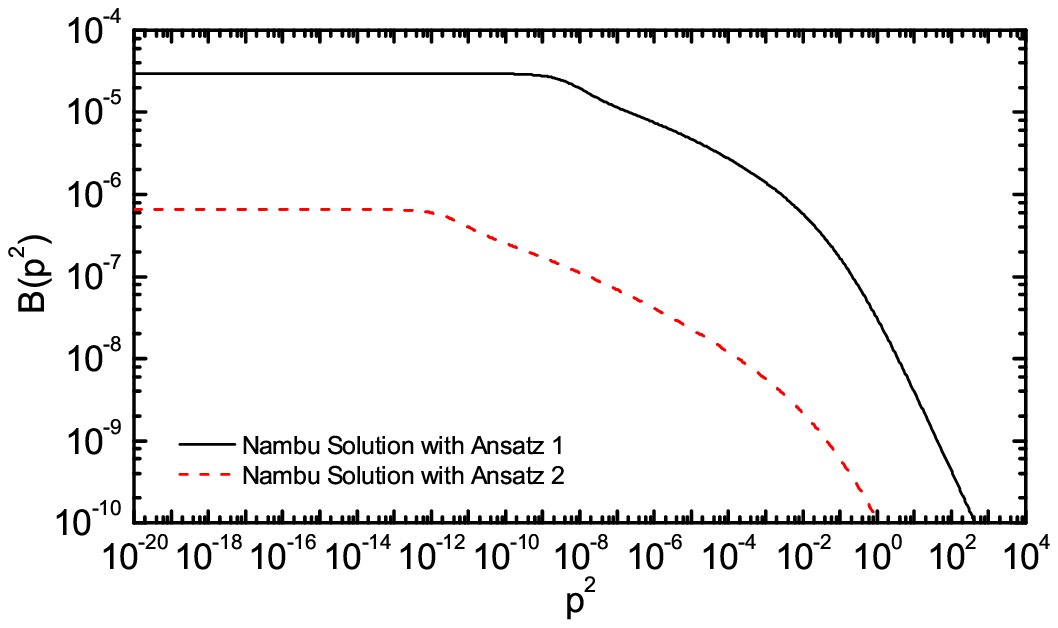}
\end{minipage}
\caption{The behavior of $B(p^{2})$ with the variation of $p^{2}$ for $\textit{N}_{f} =2$ (left) and $\textit{N}_{f} =3$ (right).}\label{fig2}
\end{figure}

\begin{figure}[h]
\begin{minipage}[t]{0.47\textwidth}
\includegraphics[width=\textwidth]{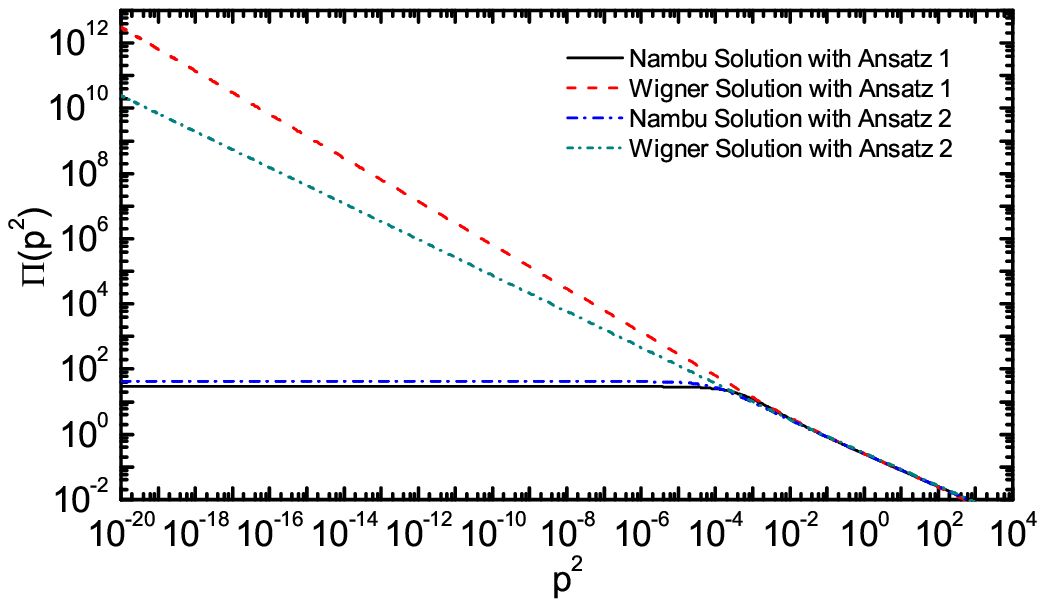}
\end{minipage}
\hfill
\begin{minipage}[t]{0.47\textwidth}
\includegraphics[width=\textwidth]{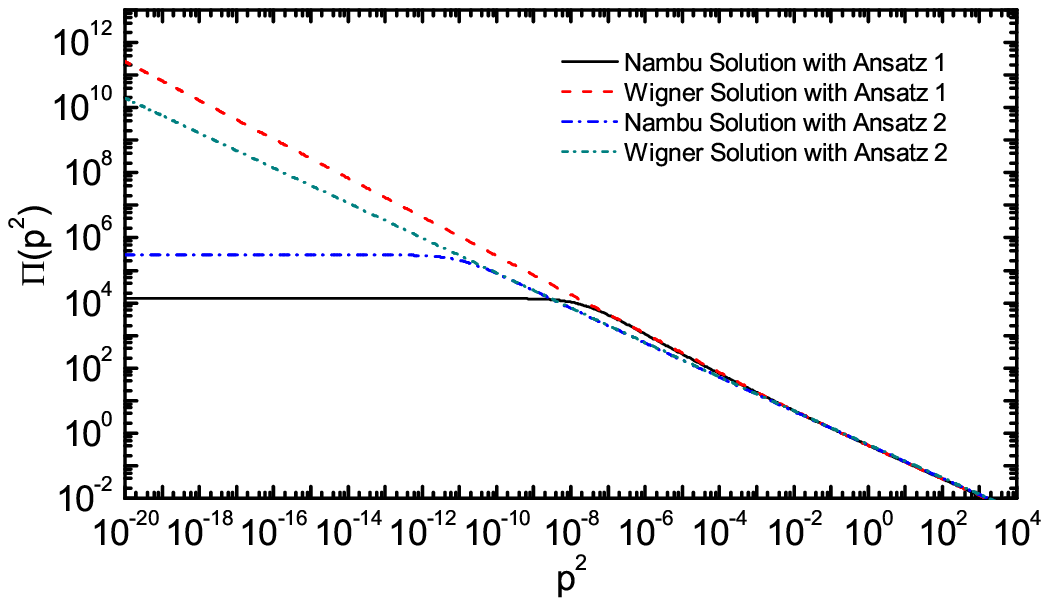}
\end{minipage}
\caption{The behavior of $\Pi(p^{2})$ with the variation of $p^{2}$ for $\textit{N}_{f} =2$ (left) and $\textit{N}_{f} =3$ (right).}\label{fig3}
\end{figure}

In order to study the phase transition between the NG phase and WN phase, one should study the different behaviors of the fermion and photon propagator in these two phases. We can numerically solve the coupled equations by the iteration method. For the NG solution, the DSEs have a non-trivial solution $B(p^{2})>0$. We start from $A(p^{2})=1$, $B(p^{2})=1$, $\Pi(p^{2})=1$ and iterate the three coupled equations until all three functions converge to a stable solution. Similarly, we can set $A(p^{2})=1$, $B(p^{2})=0$, $\Pi(p^{2})=1$ and obtain the results for $A(p^{2})$, $\Pi(p^{2})$ in the WN phase by iteration of the two coupled equations. The typical behaviors of the three functions $A(p^{2})$, $B(p^{2})$, $\Pi(p^{2})$ in the NG phase and WN phase are plotted in Figs. \ref{fig1}, Fig. \ref{fig2} and Fig. \ref{fig3} respectively.

From Fig. 1, Fig. 2 and Fig. 3 it can be seen that $A(p^{2})$, $B(p^{2})$, $\Pi(p^{2})$ in these two phases show completely different behaviors. For the NG phase, the three scalar functions are almost constant in the infrared region, and their behaviors in the ultraviolet region are: $A(p^{2})\rightarrow1$, $B(p^{2})\propto1/p^{2}$, and $\Pi(q^{2})\propto1/\sqrt{q^{2}}$, respectively. For the WN phase, in the ultraviolet region the two scalar functions coincide with their corresponding NG solutions, while in the infrared region, $A(p^{2})$ approaches zero while $\Pi(p^{2})$ tends to divergency when $p^{2}$ is close to zero, which confirms the power laws governing the infrared behavior of $QED_{3}$ in the symmetric phase (i.e., the WN phase).
\subsection{The nature of the chiral phase transition around the critical number of fermion flavors}
\begin{figure}
\includegraphics[width=0.47\textwidth]{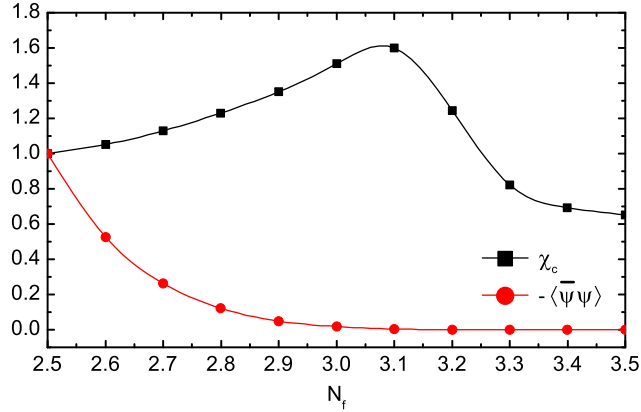}
\caption{ The dependence of -$\langle\bar\psi\psi\rangle$ and $\chi_{c}$ on fermion flavors, where parameters are normalized by their value at $\textit{N}_{f}=2.5$.}\label{fig4}
\end{figure}

Because the DSEs for the fermion and boson propagators have been reduced to three coupled equations for $A(p^{2})$, $B(p^{2})$, $\Pi(p^{2})$, we can rewrite Eq. (\ref{eq1}) in terms of $A(p^{2})$ and $B(p^{2})$
\begin{eqnarray}
\langle\bar\psi\psi\rangle= -4\int\frac{\textrm{d}^3p}{(2\pi)^3}\frac{B_{\it {N}}(p^2)}{A_{\it {N}}^{2}(p^2)p^2+B_{\it {N}}^{2}(p^2)},\label{eq24}
\end{eqnarray}
For the Ans\"{a}tze 1 of the fermion-photon vertex, substituting the NG solutions of $A(p^{2})$, $B(p^{2})$ into Eq. (\ref{eq24}) and Eq. (\ref{eq2}), we can know how the chiral condensate and chiral susceptibility change with the variation of the number of fermion flavors. The behaviors of them are plotted in Fig. \ref{fig4}.
From Fig. 4, it can be seen that with the increase of the number of fermion flavors $\textit{N}_{f}$, the chiral condensate decreases rapidly and the chiral susceptibility increases slowly. When the number of fermion flavors reaches 3.1, the chiral condensate is zero and the chiral susceptibility reaches its maximum, which shows that chiral symmetry gets restored.

\begin{figure}
\includegraphics[width=0.47\textwidth]{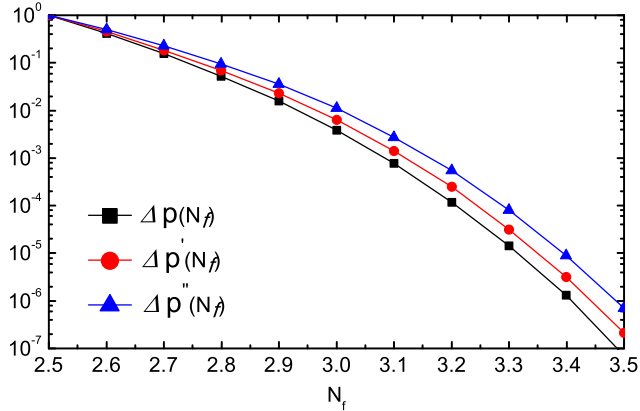}
\caption{ The dependence of $\Delta P(\textit{N}_{f})$, $\Delta P'(\textit{N}_{f})$, $\Delta P''(\textit{N}_{f})$ on fermion flavors, where parameters are normalized by their value at $\textit{N}_{f}=2.5$. }\label{fig5}
\end{figure}

In addition, we can also rewrite the Eq. (\ref{eq4}) in terms of these three functions
\begin{eqnarray}
\Delta P(N_{f})&=&-2\textit{N}_{f}\int\frac{\textrm{d}^3p}{(2\pi)^3}[\ln\frac{A_{\it {W}}^2(p^2)p^2}{A_{\it {N}}^{2}(p^2)p^2+B_{\it {N}}^{2}(p^2)}
+\frac{A_{\it {N}}(p^2)(A_{\it {N}}(p^2)-1)p^2+B_{\it {N}}^2(p^2)}{A_{\it {N}}^{2}(p^2)p^2+B_{\it {N}}^{2}(p^2)}
-\frac{A_{\it {W}}(p^2)-1}{A_{\it {W}}(p^2)}]\nonumber\\
&&+\int\frac{\textrm{d}^3p}{(2\pi)^3}[\ln\frac{1+\Pi_{\it {W}}(p^2)}{1+\Pi_{\it {N}}(p^2)}+\frac{\Pi_{\it {N}}(p^2)-\Pi_{\it {W}}(p^2)}{(1+\Pi_{\it {N}}(p^2))(1+\Pi_{\it {W}}(p^2))}],\label{eq25}
\end{eqnarray}
If we substitute different solutions of $A(p^{2})$, $B(p^{2})$, $\Pi(p^{2})$ into Eq. (\ref{eq25}) and Eq. (\ref{eq5}), we can also know how the differential pressure and its derivatives change with the variation of $\textit{N}_{f}$. Their behaviors are plotted in Fig. \ref{fig5}. 
From Fig. 5, it is found that $\Delta P(\textit{N}_{f})$, $\Delta P'(\textit{N}_{f})$ and $\Delta P''(\textit{N}_{f})$ all fall monotonically to zero as $\textit{N}_{f}$ increases and the curves show no singularity around the critical number of fermion flavors. This means that the transformation from the NG phase to the WN phase is neither of first-order nor of second-order, but may be a high-order continuous phase transition, when the number of fermion flavors reaches the critical value.

\begin{figure}
\includegraphics[width=0.47\textwidth]{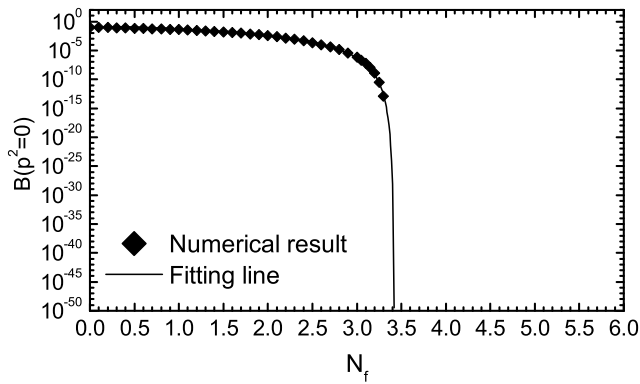}
\caption{The infrared values of the self-energy function $B(p^{2})$ as a function of the number of fermion flavors $N_{f}$. }\label{fig6}
\end{figure}

In order to consider the impact of a non-trivial extension of the bare vertex approximation on the nature of chiral phase transitions, here we adopt Ans\"{a}tze 2 of the fermion每photon vertex (the BC simplified vertex) to study this problem. It is well known that the CJT effective potential adopted in our manuscript corresponds to the bare vertex approximation for solving the fermion and the boson propagator. If one tries to go beyond the bare vertex approximation (for example, the BC-simplified
vertex), up to now people do not know how to construct the corresponding effective potential like CJT. Therefore, we cannot use the differential pressure between NG phase and WN phase to study the phase transition. In this circumstance, a method based on the chiral susceptibility can work, as explained in Ref.~\cite{PhysRevLett.106.172301}. In this work we will employ a different order parameter instead, namely, the infrared values of self-energy function $B(p^2=0)$~\cite{ProgPartNuclPhys.45.S1,PhysRevD.82.034006}, of which a nonzero value implies a nonzero condensate. The analytical expression for this quantity is investigated and given by T.W. Appelquist \textit{et al.}~\cite{PhysRevD.33.3704}
\begin{eqnarray}
B(0)=aN_{f}\exp^{-2\pi/\sqrt{N_{f,c}/N_{f}-1}} ,\label{eq26}
\end{eqnarray}
where the notations $a$ and $N_{f,c}$ are the fitting parameters. For our numerical result, it can be found that the best fitting parameters are $(a, N_{f,c})=(2.96, 3.43)$. The numerical result together with the fitting line are plotted in Fig. \ref{fig6}. 
 We can see clearly that the infrared value of $B(p^{2})$ decreases with the fermion flavors increasing. When the number of fermion flavors approaches the critical value 3.43, the infrared value of $B(p^{2})$ drops rapidly, which signals a typical characteristic of second-order phase transition.

\section{summary and conclusions}
In this paper, we discuss the nature of the chiral phase transition around the critical number of fermion flavors in QED$_3$ with $\textit{N}_{f}$ massless fermion flavors at zero temperature and density. For this, we firstly introduce the criteria determining the locations and characteristics of the chiral phase transition, and then in the Landau gauge, we numerically solve the coupled Dyson每Schwinger equations within the bare and simplified BC vertices separately. The numerical results show that, in the bare vertex approximation, the system undergoes a high-order continuous phase transition from the NG phase into the WN phase when the number of fermion flavors $\textit{N}_{f}$ reaches the critical number $\textit{N}_{f,c}$. Finally, in order to analyze the impacts of different choices of vertex Ans\"{a}tze on the results of our paper, we have also adopted the simplified BC vertex to study this problem. As is shown in Fig. 6, for the simplified BC vertex the chiral phase transition is a typical second-order phase transition. This shows that the impacts of the different choices of vertex Ans$\rm{{\ddot a}}$tze are important for the study of chiral phase transition of QED$_3$. Undoubtedly this problem deserves further investigation.
\bigskip
\acknowledgments
This work is supported in part by the National Natural Science Foundation of China (under Grants 11275097, 10935001, 11274166 and 11075075), the National Basic Research Program of China (under Grant 2012CB921504) and the Research Fund for the Doctoral Program of Higher Education (under Grant No 2012009111002).
\bibliographystyle{apsrev4-1}
\bibliography{AOP69451}

\begin{thebibliography}{33}%
\makeatletter
\providecommand \@ifxundefined [1]{%
 \@ifx{#1\undefined}
}%
\providecommand \@ifnum [1]{%
 \ifnum #1\expandafter \@firstoftwo
 \else \expandafter \@secondoftwo
 \fi
}%
\providecommand \@ifx [1]{%
 \ifx #1\expandafter \@firstoftwo
 \else \expandafter \@secondoftwo
 \fi
}%
\providecommand \natexlab [1]{#1}%
\providecommand \enquote  [1]{``#1''}%
\providecommand \bibnamefont  [1]{#1}%
\providecommand \bibfnamefont [1]{#1}%
\providecommand \citenamefont [1]{#1}%
\providecommand \href@noop [0]{\@secondoftwo}%
\providecommand \href [0]{\begingroup \@sanitize@url \@href}%
\providecommand \@href[1]{\@@startlink{#1}\@@href}%
\providecommand \@@href[1]{\endgroup#1\@@endlink}%
\providecommand \@sanitize@url [0]{\catcode `\\12\catcode `\$12\catcode
  `\&12\catcode `\#12\catcode `\^12\catcode `\_12\catcode `\%12\relax}%
\providecommand \@@startlink[1]{}%
\providecommand \@@endlink[0]{}%
\providecommand \url  [0]{\begingroup\@sanitize@url \@url }%
\providecommand \@url [1]{\endgroup\@href {#1}{\urlprefix }}%
\providecommand \urlprefix  [0]{URL }%
\providecommand \Eprint [0]{\href }%
\providecommand \doibase [0]{http://dx.doi.org/}%
\providecommand \selectlanguage [0]{\@gobble}%
\providecommand \bibinfo  [0]{\@secondoftwo}%
\providecommand \bibfield  [0]{\@secondoftwo}%
\providecommand \translation [1]{[#1]}%
\providecommand \BibitemOpen [0]{}%
\providecommand \bibitemStop [0]{}%
\providecommand \bibitemNoStop [0]{.\EOS\space}%
\providecommand \EOS [0]{\spacefactor3000\relax}%
\providecommand \BibitemShut  [1]{\csname bibitem#1\endcsname}%
\let\auto@bib@innerbib\@empty
\bibitem [{\citenamefont {Pisarski}(1984)}]{PhysRevD.29.2423}%
  \BibitemOpen
  \bibfield  {author} {\bibinfo {author} {\bibfnamefont {R.~D.}\ \bibnamefont
  {Pisarski}},\ }\href {\doibase 10.1103/PhysRevD.29.2423} {\bibfield
  {journal} {\bibinfo  {journal} {Phys. Rev. D}\ }\textbf {\bibinfo {volume}
  {29}},\ \bibinfo {pages} {2423} (\bibinfo {year} {1984})}\BibitemShut
  {NoStop}%
\bibitem [{\citenamefont {Appelquist}\ \emph {et~al.}(1985)\citenamefont
  {Appelquist}, \citenamefont {Bowick}, \citenamefont {Cohler},\ and\
  \citenamefont {Wijewardhana}}]{PhysRevLett.55.1715}%
  \BibitemOpen
  \bibfield  {author} {\bibinfo {author} {\bibfnamefont {T.}~\bibnamefont
  {Appelquist}}, \bibinfo {author} {\bibfnamefont {M.~J.}\ \bibnamefont
  {Bowick}}, \bibinfo {author} {\bibfnamefont {E.}~\bibnamefont {Cohler}}, \
  and\ \bibinfo {author} {\bibfnamefont {L.~C.~R.}\ \bibnamefont
  {Wijewardhana}},\ }\href {\doibase 10.1103/PhysRevLett.55.1715} {\bibfield
  {journal} {\bibinfo  {journal} {Phys. Rev. Lett.}\ }\textbf {\bibinfo
  {volume} {55}},\ \bibinfo {pages} {1715} (\bibinfo {year}
  {1985})}\BibitemShut {NoStop}%
\bibitem [{\citenamefont {Appelquist}\ \emph {et~al.}(1988)\citenamefont
  {Appelquist}, \citenamefont {Nash},\ and\ \citenamefont
  {Wijewardhana}}]{PhysRevLett.60.2575}%
  \BibitemOpen
  \bibfield  {author} {\bibinfo {author} {\bibfnamefont {T.}~\bibnamefont
  {Appelquist}}, \bibinfo {author} {\bibfnamefont {D.}~\bibnamefont {Nash}}, \
  and\ \bibinfo {author} {\bibfnamefont {L.~C.~R.}\ \bibnamefont
  {Wijewardhana}},\ }\href {\doibase 10.1103/PhysRevLett.60.2575} {\bibfield
  {journal} {\bibinfo  {journal} {Phys. Rev. Lett.}\ }\textbf {\bibinfo
  {volume} {60}},\ \bibinfo {pages} {2575} (\bibinfo {year}
  {1988})}\BibitemShut {NoStop}%
\bibitem [{\citenamefont {Nash}(1989)}]{PhysRevLett.62.3024}%
  \BibitemOpen
  \bibfield  {author} {\bibinfo {author} {\bibfnamefont {D.}~\bibnamefont
  {Nash}},\ }\href {\doibase 10.1103/PhysRevLett.62.3024} {\bibfield  {journal}
  {\bibinfo  {journal} {Phys. Rev. Lett.}\ }\textbf {\bibinfo {volume} {62}},\
  \bibinfo {pages} {3024} (\bibinfo {year} {1989})}\BibitemShut {NoStop}%
\bibitem [{\citenamefont {Burden}\ and\ \citenamefont
  {Roberts}(1991)}]{PhysRevD.44.540}%
  \BibitemOpen
  \bibfield  {author} {\bibinfo {author} {\bibfnamefont {C.~J.}\ \bibnamefont
  {Burden}}\ and\ \bibinfo {author} {\bibfnamefont {C.~D.}\ \bibnamefont
  {Roberts}},\ }\href {\doibase 10.1103/PhysRevD.44.540} {\bibfield  {journal}
  {\bibinfo  {journal} {Phys. Rev. D}\ }\textbf {\bibinfo {volume} {44}},\
  \bibinfo {pages} {540} (\bibinfo {year} {1991})}\BibitemShut {NoStop}%
\bibitem [{\citenamefont {Curtis}\ \emph {et~al.}(1992)\citenamefont {Curtis},
  \citenamefont {Pennington},\ and\ \citenamefont {Walsh}}]{PhysLettB.295.313}%
  \BibitemOpen
  \bibfield  {author} {\bibinfo {author} {\bibfnamefont {D.}~\bibnamefont
  {Curtis}}, \bibinfo {author} {\bibfnamefont {M.}~\bibnamefont {Pennington}},
  \ and\ \bibinfo {author} {\bibfnamefont {D.}~\bibnamefont {Walsh}},\ }\href
  {\doibase http://dx.doi.org/10.1016/0370-2693(92)91572-Q} {\bibfield
  {journal} {\bibinfo  {journal} {Phys. Lett. B}\ }\textbf {\bibinfo {volume}
  {295}},\ \bibinfo {pages} {313 } (\bibinfo {year} {1992})}\BibitemShut
  {NoStop}%
\bibitem [{\citenamefont {Kondo}\ and\ \citenamefont
  {Maris}(1995)}]{PhysRevLett.74.18}%
  \BibitemOpen
  \bibfield  {author} {\bibinfo {author} {\bibfnamefont {K.~I.}\ \bibnamefont
  {Kondo}}\ and\ \bibinfo {author} {\bibfnamefont {P.}~\bibnamefont {Maris}},\
  }\href {\doibase 10.1103/PhysRevLett.74.18} {\bibfield  {journal} {\bibinfo
  {journal} {Phys. Rev. Lett.}\ }\textbf {\bibinfo {volume} {74}},\ \bibinfo
  {pages} {18} (\bibinfo {year} {1995})}\BibitemShut {NoStop}%
\bibitem [{\citenamefont {Maris}(1996)}]{PhysRevD.54.4049}%
  \BibitemOpen
  \bibfield  {author} {\bibinfo {author} {\bibfnamefont {P.}~\bibnamefont
  {Maris}},\ }\href {\doibase 10.1103/PhysRevD.54.4049} {\bibfield  {journal}
  {\bibinfo  {journal} {Phys. Rev. D}\ }\textbf {\bibinfo {volume} {54}},\
  \bibinfo {pages} {4049} (\bibinfo {year} {1996})}\BibitemShut {NoStop}%
\bibitem [{\citenamefont {Bashir}(2000)}]{PhysLettB.491.280}%
  \BibitemOpen
  \bibfield  {author} {\bibinfo {author} {\bibfnamefont {A.}~\bibnamefont
  {Bashir}},\ }\href {\doibase http://dx.doi.org/10.1016/S0370-2693(00)01043-1}
  {\bibfield  {journal} {\bibinfo  {journal} {Phys. Lett. B}\ }\textbf
  {\bibinfo {volume} {491}},\ \bibinfo {pages} {280 } (\bibinfo {year}
  {2000})}\BibitemShut {NoStop}%
\bibitem [{\citenamefont {Gusynin}\ and\ \citenamefont
  {Reenders}(2003)}]{PhysRevD.68.025017}%
  \BibitemOpen
  \bibfield  {author} {\bibinfo {author} {\bibfnamefont {V.~P.}\ \bibnamefont
  {Gusynin}}\ and\ \bibinfo {author} {\bibfnamefont {M.}~\bibnamefont
  {Reenders}},\ }\href {\doibase 10.1103/PhysRevD.68.025017} {\bibfield
  {journal} {\bibinfo  {journal} {Phys. Rev. D}\ }\textbf {\bibinfo {volume}
  {68}},\ \bibinfo {pages} {025017} (\bibinfo {year} {2003})}\BibitemShut
  {NoStop}%
\bibitem [{\citenamefont {Feng}\ \emph {et~al.}(2006)\citenamefont {Feng},
  \citenamefont {Hou}, \citenamefont {He}, \citenamefont {Sun},\ and\
  \citenamefont {Zong}}]{PhysRevD.73.016004}%
  \BibitemOpen
  \bibfield  {author} {\bibinfo {author} {\bibfnamefont {H.-t.}\ \bibnamefont
  {Feng}}, \bibinfo {author} {\bibfnamefont {F.-y.}\ \bibnamefont {Hou}},
  \bibinfo {author} {\bibfnamefont {X.}~\bibnamefont {He}}, \bibinfo {author}
  {\bibfnamefont {W.-m.}\ \bibnamefont {Sun}}, \ and\ \bibinfo {author}
  {\bibfnamefont {H.-s.}\ \bibnamefont {Zong}},\ }\href {\doibase
  10.1103/PhysRevD.73.016004} {\bibfield  {journal} {\bibinfo  {journal} {Phys.
  Rev. D}\ }\textbf {\bibinfo {volume} {73}},\ \bibinfo {pages} {016004}
  (\bibinfo {year} {2006})}\BibitemShut {NoStop}%
\bibitem [{\citenamefont {Bashir}\ \emph {et~al.}(2008)\citenamefont {Bashir},
  \citenamefont {Raya}, \citenamefont {Clo\"et},\ and\ \citenamefont
  {Roberts}}]{PhysRevC.78.055201}%
  \BibitemOpen
  \bibfield  {author} {\bibinfo {author} {\bibfnamefont {A.}~\bibnamefont
  {Bashir}}, \bibinfo {author} {\bibfnamefont {A.}~\bibnamefont {Raya}},
  \bibinfo {author} {\bibfnamefont {I.~C.}\ \bibnamefont {Clo\"et}}, \ and\
  \bibinfo {author} {\bibfnamefont {C.~D.}\ \bibnamefont {Roberts}},\ }\href
  {\doibase 10.1103/PhysRevC.78.055201} {\bibfield  {journal} {\bibinfo
  {journal} {Phys. Rev. C}\ }\textbf {\bibinfo {volume} {78}},\ \bibinfo
  {pages} {055201} (\bibinfo {year} {2008})}\BibitemShut {NoStop}%
\bibitem [{\citenamefont {Bashir}\ \emph {et~al.}(2011)\citenamefont {Bashir},
  \citenamefont {Raya},\ and\ \citenamefont
  {S\'anchez-Madrigal}}]{PhysRevD.84.036013}%
  \BibitemOpen
  \bibfield  {author} {\bibinfo {author} {\bibfnamefont {A.}~\bibnamefont
  {Bashir}}, \bibinfo {author} {\bibfnamefont {A.}~\bibnamefont {Raya}}, \ and\
  \bibinfo {author} {\bibfnamefont {S.}~\bibnamefont {S\'anchez-Madrigal}},\
  }\href {\doibase 10.1103/PhysRevD.84.036013} {\bibfield  {journal} {\bibinfo
  {journal} {Phys. Rev. D}\ }\textbf {\bibinfo {volume} {84}},\ \bibinfo
  {pages} {036013} (\bibinfo {year} {2011})}\BibitemShut {NoStop}%
\bibitem [{\citenamefont {Feng}\ \emph {et~al.}(2012)\citenamefont {Feng},
  \citenamefont {Wang}, \citenamefont {Sun},\ and\ \citenamefont
  {Zong}}]{PhysRevD.86.105042}%
  \BibitemOpen
  \bibfield  {author} {\bibinfo {author} {\bibfnamefont {H.-t.}\ \bibnamefont
  {Feng}}, \bibinfo {author} {\bibfnamefont {B.}~\bibnamefont {Wang}}, \bibinfo
  {author} {\bibfnamefont {W.-m.}\ \bibnamefont {Sun}}, \ and\ \bibinfo
  {author} {\bibfnamefont {H.-s.}\ \bibnamefont {Zong}},\ }\href {\doibase
  10.1103/PhysRevD.86.105042} {\bibfield  {journal} {\bibinfo  {journal} {Phys.
  Rev. D}\ }\textbf {\bibinfo {volume} {86}},\ \bibinfo {pages} {105042}
  (\bibinfo {year} {2012})}\BibitemShut {NoStop}%
\bibitem [{\citenamefont {G{\"o}pfert}\ and\ \citenamefont
  {Mack}(1982)}]{CommunMathPhys.82.545}%
  \BibitemOpen
  \bibfield  {author} {\bibinfo {author} {\bibfnamefont {M.}~\bibnamefont
  {G{\"o}pfert}}\ and\ \bibinfo {author} {\bibfnamefont {G.}~\bibnamefont
  {Mack}},\ }\href {\doibase 10.1007/BF01961240} {\bibfield  {journal}
  {\bibinfo  {journal} {Commun. Math. Phys.}\ }\textbf {\bibinfo {volume}
  {82}},\ \bibinfo {pages} {545} (\bibinfo {year} {1982})}\BibitemShut
  {NoStop}%
\bibitem [{\citenamefont {Burden}\ \emph {et~al.}(1992)\citenamefont {Burden},
  \citenamefont {Praschifka},\ and\ \citenamefont
  {Roberts}}]{PhysRevD.46.2695}%
  \BibitemOpen
  \bibfield  {author} {\bibinfo {author} {\bibfnamefont {C.~J.}\ \bibnamefont
  {Burden}}, \bibinfo {author} {\bibfnamefont {J.}~\bibnamefont {Praschifka}},
  \ and\ \bibinfo {author} {\bibfnamefont {C.~D.}\ \bibnamefont {Roberts}},\
  }\href {\doibase 10.1103/PhysRevD.46.2695} {\bibfield  {journal} {\bibinfo
  {journal} {Phys. Rev. D}\ }\textbf {\bibinfo {volume} {46}},\ \bibinfo
  {pages} {2695} (\bibinfo {year} {1992})}\BibitemShut {NoStop}%
\bibitem [{\citenamefont {Maris}(1995)}]{PhysRevD.52.6087}%
  \BibitemOpen
  \bibfield  {author} {\bibinfo {author} {\bibfnamefont {P.}~\bibnamefont
  {Maris}},\ }\href {\doibase 10.1103/PhysRevD.52.6087} {\bibfield  {journal}
  {\bibinfo  {journal} {Phys. Rev. D}\ }\textbf {\bibinfo {volume} {52}},\
  \bibinfo {pages} {6087} (\bibinfo {year} {1995})}\BibitemShut {NoStop}%
\bibitem [{\citenamefont {Herbut}\ and\ \citenamefont
  {Seradjeh}(2003)}]{PhysRevLett.91.171601}%
  \BibitemOpen
  \bibfield  {author} {\bibinfo {author} {\bibfnamefont {I.~F.}\ \bibnamefont
  {Herbut}}\ and\ \bibinfo {author} {\bibfnamefont {B.~H.}\ \bibnamefont
  {Seradjeh}},\ }\href {\doibase 10.1103/PhysRevLett.91.171601} {\bibfield
  {journal} {\bibinfo  {journal} {Phys. Rev. Lett.}\ }\textbf {\bibinfo
  {volume} {91}},\ \bibinfo {pages} {171601} (\bibinfo {year}
  {2003})}\BibitemShut {NoStop}%
\bibitem [{\citenamefont {Rantner}\ and\ \citenamefont
  {Wen}(2001)}]{PhysRevLett.86.3871}%
  \BibitemOpen
  \bibfield  {author} {\bibinfo {author} {\bibfnamefont {W.}~\bibnamefont
  {Rantner}}\ and\ \bibinfo {author} {\bibfnamefont {X.-g.}\ \bibnamefont
  {Wen}},\ }\href {\doibase 10.1103/PhysRevLett.86.3871} {\bibfield  {journal}
  {\bibinfo  {journal} {Phys. Rev. Lett.}\ }\textbf {\bibinfo {volume} {86}},\
  \bibinfo {pages} {3871} (\bibinfo {year} {2001})}\BibitemShut {NoStop}%
\bibitem [{\citenamefont {Franz}\ \emph {et~al.}(2002)\citenamefont {Franz},
  \citenamefont {Te\ifmmode \check{s}\else
  \v{s}\fi{}anovi\ifmmode~\acute{c}\else \'{c}\fi{}},\ and\ \citenamefont
  {Vafek}}]{PhysRevB.66.054535}%
  \BibitemOpen
  \bibfield  {author} {\bibinfo {author} {\bibfnamefont {M.}~\bibnamefont
  {Franz}}, \bibinfo {author} {\bibfnamefont {Z.}~\bibnamefont {Te\ifmmode
  \check{s}\else \v{s}\fi{}anovi\ifmmode~\acute{c}\else \'{c}\fi{}}}, \ and\
  \bibinfo {author} {\bibfnamefont {O.}~\bibnamefont {Vafek}},\ }\href
  {\doibase 10.1103/PhysRevB.66.054535} {\bibfield  {journal} {\bibinfo
  {journal} {Phys. Rev. B}\ }\textbf {\bibinfo {volume} {66}},\ \bibinfo
  {pages} {054535} (\bibinfo {year} {2002})}\BibitemShut {NoStop}%
\bibitem [{\citenamefont {Liu}(2005)}]{PhysRevB.71.172501}%
  \BibitemOpen
  \bibfield  {author} {\bibinfo {author} {\bibfnamefont {G.-z.}\ \bibnamefont
  {Liu}},\ }\href {\doibase 10.1103/PhysRevB.71.172501} {\bibfield  {journal}
  {\bibinfo  {journal} {Phys. Rev. B}\ }\textbf {\bibinfo {volume} {71}},\
  \bibinfo {pages} {172501} (\bibinfo {year} {2005})}\BibitemShut {NoStop}%
\bibitem [{\citenamefont {Lee}\ \emph {et~al.}(2006)\citenamefont {Lee},
  \citenamefont {Nagaosa},\ and\ \citenamefont {Wen}}]{RevModPhys.78.17}%
  \BibitemOpen
  \bibfield  {author} {\bibinfo {author} {\bibfnamefont {P.~A.}\ \bibnamefont
  {Lee}}, \bibinfo {author} {\bibfnamefont {N.}~\bibnamefont {Nagaosa}}, \ and\
  \bibinfo {author} {\bibfnamefont {X.-g.}\ \bibnamefont {Wen}},\ }\href
  {\doibase 10.1103/RevModPhys.78.17} {\bibfield  {journal} {\bibinfo
  {journal} {Rev. Mod. Phys.}\ }\textbf {\bibinfo {volume} {78}},\ \bibinfo
  {pages} {17} (\bibinfo {year} {2006})}\BibitemShut {NoStop}%
\bibitem [{\citenamefont {Drut}\ and\ \citenamefont
  {L\"ahde}(2009)}]{PhysRevLett.102.026802}%
  \BibitemOpen
  \bibfield  {author} {\bibinfo {author} {\bibfnamefont {J.~E.}\ \bibnamefont
  {Drut}}\ and\ \bibinfo {author} {\bibfnamefont {T.~A.}\ \bibnamefont
  {L\"ahde}},\ }\href {\doibase 10.1103/PhysRevLett.102.026802} {\bibfield
  {journal} {\bibinfo  {journal} {Phys. Rev. Lett.}\ }\textbf {\bibinfo
  {volume} {102}},\ \bibinfo {pages} {026802} (\bibinfo {year}
  {2009})}\BibitemShut {NoStop}%
\bibitem [{\citenamefont {Zhang}\ \emph {et~al.}(2011)\citenamefont {Zhang},
  \citenamefont {Liu},\ and\ \citenamefont {Huang}}]{PhysRevB.83.115438}%
  \BibitemOpen
  \bibfield  {author} {\bibinfo {author} {\bibfnamefont {C.-x.}\ \bibnamefont
  {Zhang}}, \bibinfo {author} {\bibfnamefont {G.-z.}\ \bibnamefont {Liu}}, \
  and\ \bibinfo {author} {\bibfnamefont {M.-q.}\ \bibnamefont {Huang}},\ }\href
  {\doibase 10.1103/PhysRevB.83.115438} {\bibfield  {journal} {\bibinfo
  {journal} {Phys. Rev. B}\ }\textbf {\bibinfo {volume} {83}},\ \bibinfo
  {pages} {115438} (\bibinfo {year} {2011})}\BibitemShut {NoStop}%
\bibitem [{\citenamefont {Aoki}\ \emph {et~al.}(2006)\citenamefont {Aoki},
  \citenamefont {Endr{\H{o}}di}, \citenamefont {Fodor}, \citenamefont {Katz},\
  and\ \citenamefont {Szabo}}]{Nature.443.675}%
  \BibitemOpen
  \bibfield  {author} {\bibinfo {author} {\bibfnamefont {Y.}~\bibnamefont
  {Aoki}}, \bibinfo {author} {\bibfnamefont {G.}~\bibnamefont {Endr{\H{o}}di}},
  \bibinfo {author} {\bibfnamefont {Z.}~\bibnamefont {Fodor}}, \bibinfo
  {author} {\bibfnamefont {S.}~\bibnamefont {Katz}}, \ and\ \bibinfo {author}
  {\bibfnamefont {K.}~\bibnamefont {Szabo}},\ }\href@noop {} {\bibfield
  {journal} {\bibinfo  {journal} {Nature}\ }\textbf {\bibinfo {volume} {443}},\
  \bibinfo {pages} {675} (\bibinfo {year} {2006})}\BibitemShut {NoStop}%
\bibitem [{\citenamefont {Contrera}\ \emph {et~al.}(2010)\citenamefont
  {Contrera}, \citenamefont {Orsaria},\ and\ \citenamefont
  {Scoccola}}]{PhysRevD.82.054026}%
  \BibitemOpen
  \bibfield  {author} {\bibinfo {author} {\bibfnamefont {G.~A.}\ \bibnamefont
  {Contrera}}, \bibinfo {author} {\bibfnamefont {M.}~\bibnamefont {Orsaria}}, \
  and\ \bibinfo {author} {\bibfnamefont {N.~N.}\ \bibnamefont {Scoccola}},\
  }\href {\doibase 10.1103/PhysRevD.82.054026} {\bibfield  {journal} {\bibinfo
  {journal} {Phys. Rev. D}\ }\textbf {\bibinfo {volume} {82}},\ \bibinfo
  {pages} {054026} (\bibinfo {year} {2010})}\BibitemShut {NoStop}%
\bibitem [{\citenamefont {Cornwall}\ \emph {et~al.}(1974)\citenamefont
  {Cornwall}, \citenamefont {Jackiw},\ and\ \citenamefont
  {Tomboulis}}]{PhysRevD.10.2428}%
  \BibitemOpen
  \bibfield  {author} {\bibinfo {author} {\bibfnamefont {J.~M.}\ \bibnamefont
  {Cornwall}}, \bibinfo {author} {\bibfnamefont {R.}~\bibnamefont {Jackiw}}, \
  and\ \bibinfo {author} {\bibfnamefont {E.}~\bibnamefont {Tomboulis}},\ }\href
  {\doibase 10.1103/PhysRevD.10.2428} {\bibfield  {journal} {\bibinfo
  {journal} {Phys. Rev. D}\ }\textbf {\bibinfo {volume} {10}},\ \bibinfo
  {pages} {2428} (\bibinfo {year} {1974})}\BibitemShut {NoStop}%
\bibitem [{\citenamefont {Bashir}\ \emph {et~al.}(2012)\citenamefont {Bashir},
  \citenamefont {Bermudez}, \citenamefont {Chang},\ and\ \citenamefont
  {Roberts}}]{PhysRevC.85.045205}%
  \BibitemOpen
  \bibfield  {author} {\bibinfo {author} {\bibfnamefont {A.}~\bibnamefont
  {Bashir}}, \bibinfo {author} {\bibfnamefont {R.}~\bibnamefont {Bermudez}},
  \bibinfo {author} {\bibfnamefont {L.}~\bibnamefont {Chang}}, \ and\ \bibinfo
  {author} {\bibfnamefont {C.~D.}\ \bibnamefont {Roberts}},\ }\href {\doibase
  10.1103/PhysRevC.85.045205} {\bibfield  {journal} {\bibinfo  {journal} {Phys.
  Rev. C}\ }\textbf {\bibinfo {volume} {85}},\ \bibinfo {pages} {045205}
  (\bibinfo {year} {2012})}\BibitemShut {NoStop}%
\bibitem [{\citenamefont {Qin}\ \emph {et~al.}(2013)\citenamefont {Qin},
  \citenamefont {Chang}, \citenamefont {Liu}, \citenamefont {Roberts},\ and\
  \citenamefont {Schmidt}}]{PhysLettB.722.384}%
  \BibitemOpen
  \bibfield  {author} {\bibinfo {author} {\bibfnamefont {S.-x.}\ \bibnamefont
  {Qin}}, \bibinfo {author} {\bibfnamefont {L.}~\bibnamefont {Chang}}, \bibinfo
  {author} {\bibfnamefont {Y.-x.}\ \bibnamefont {Liu}}, \bibinfo {author}
  {\bibfnamefont {C.~D.}\ \bibnamefont {Roberts}}, \ and\ \bibinfo {author}
  {\bibfnamefont {S.~M.}\ \bibnamefont {Schmidt}},\ }\href {\doibase
  http://dx.doi.org/10.1016/j.physletb.2013.04.034} {\bibfield  {journal}
  {\bibinfo  {journal} {Phys. Lett. B}\ }\textbf {\bibinfo {volume} {722}},\
  \bibinfo {pages} {384 } (\bibinfo {year} {2013})}\BibitemShut {NoStop}%
\bibitem [{\citenamefont {Qin}\ \emph {et~al.}(2011)\citenamefont {Qin},
  \citenamefont {Chang}, \citenamefont {Chen}, \citenamefont {Liu},\ and\
  \citenamefont {Roberts}}]{PhysRevLett.106.172301}%
  \BibitemOpen
  \bibfield  {author} {\bibinfo {author} {\bibfnamefont {S.-x.}\ \bibnamefont
  {Qin}}, \bibinfo {author} {\bibfnamefont {L.}~\bibnamefont {Chang}}, \bibinfo
  {author} {\bibfnamefont {H.}~\bibnamefont {Chen}}, \bibinfo {author}
  {\bibfnamefont {Y.-x.}\ \bibnamefont {Liu}}, \ and\ \bibinfo {author}
  {\bibfnamefont {C.~D.}\ \bibnamefont {Roberts}},\ }\href {\doibase
  10.1103/PhysRevLett.106.172301} {\bibfield  {journal} {\bibinfo  {journal}
  {Phys. Rev. Lett.}\ }\textbf {\bibinfo {volume} {106}},\ \bibinfo {pages}
  {172301} (\bibinfo {year} {2011})}\BibitemShut {NoStop}%
\bibitem [{\citenamefont {Roberts}\ and\ \citenamefont
  {Schmidt}(2000)}]{ProgPartNuclPhys.45.S1}%
  \BibitemOpen
  \bibfield  {author} {\bibinfo {author} {\bibfnamefont {C.}~\bibnamefont
  {Roberts}}\ and\ \bibinfo {author} {\bibfnamefont {S.}~\bibnamefont
  {Schmidt}},\ }\href {\doibase
  http://dx.doi.org/10.1016/S0146-6410(00)90011-5} {\bibfield  {journal}
  {\bibinfo  {journal} {Prog. Part. Nucl. Phys.}\ }\textbf {\bibinfo {volume}
  {45}},\ \bibinfo {pages} {S1 } (\bibinfo {year} {2000})}\BibitemShut
  {NoStop}%
\bibitem [{\citenamefont {Blank}\ and\ \citenamefont
  {Krassnigg}(2010)}]{PhysRevD.82.034006}%
  \BibitemOpen
  \bibfield  {author} {\bibinfo {author} {\bibfnamefont {M.}~\bibnamefont
  {Blank}}\ and\ \bibinfo {author} {\bibfnamefont {A.}~\bibnamefont
  {Krassnigg}},\ }\href {\doibase 10.1103/PhysRevD.82.034006} {\bibfield
  {journal} {\bibinfo  {journal} {Phys. Rev. D}\ }\textbf {\bibinfo {volume}
  {82}},\ \bibinfo {pages} {034006} (\bibinfo {year} {2010})}\BibitemShut
  {NoStop}%
\bibitem [{\citenamefont {Appelquist}\ \emph {et~al.}(1986)\citenamefont
  {Appelquist}, \citenamefont {Bowick}, \citenamefont {Karabali},\ and\
  \citenamefont {Wijewardhana}}]{PhysRevD.33.3704}%
  \BibitemOpen
  \bibfield  {author} {\bibinfo {author} {\bibfnamefont {T.~W.}\ \bibnamefont
  {Appelquist}}, \bibinfo {author} {\bibfnamefont {M.}~\bibnamefont {Bowick}},
  \bibinfo {author} {\bibfnamefont {D.}~\bibnamefont {Karabali}}, \ and\
  \bibinfo {author} {\bibfnamefont {L.~C.~R.}\ \bibnamefont {Wijewardhana}},\
  }\href {\doibase 10.1103/PhysRevD.33.3704} {\bibfield  {journal} {\bibinfo
  {journal} {Phys. Rev. D}\ }\textbf {\bibinfo {volume} {33}},\ \bibinfo
  {pages} {3704} (\bibinfo {year} {1986})}\BibitemShut {NoStop}%
\end{thebibliography}%

\end{document}